\documentclass[conference]{IEEEtran}
\usepackage{psfrag}

\usepackage{cite}

\ifCLASSINFOpdf
   \usepackage[pdftex]{graphicx}
\else
   \usepackage[dvips]{graphicx}
\fi
\usepackage{amssymb}
\usepackage[cmex10]{amsmath}
\usepackage{algorithm}
\usepackage{algorithmic}

\begin{document}
%
\title{Exploiting Frequency and Spatial Dimensions in Small Cell Wireless Networks}
\author{\IEEEauthorblockN{Stelios~Stefanatos~and~Angeliki~Alexiou,}
\IEEEauthorblockA{Department~of~Digital~Systems,~University~of~Piraeus,~Greece}}


%


\maketitle

\begin{abstract}
This paper examines the efficiency of spatial and frequency dimensions in serving multiple users in the downlink of a small cell wireless network with randomly deployed access points. For this purpose, the stochastic geometry framework is incorporated,  taking into account the user distribution within each cell and the effect of sharing the available system resources to multiple users. An analysis of performance in terms of signal-to-interference-ratio and achieved user rate is provided that holds under the class of non-cooperative multiple access schemes. In order to obtain concrete results, two simple instances of multiple access schemes are considered. It is shown that performance depends critically on both the availability of frequency and/or spatial dimensions as well as the way they are employed. In particular, increasing the number of available frequency dimensions alone is beneficial for users experiencing large interference, whereas increasing spatial dimensions without employing frequency dimensions degrades performance. However, best performance is achieved when both dimensions are combined in serving the users.

\end{abstract}


%
\IEEEpeerreviewmaketitle

\section{Introduction}
Small cell networks have attracted a lot of attention recently as they are considered a promising method to satisfy the ever increasing rate demands of wireless users. These networks are based on densely-deployed, low-cost access points (APs) whose density $\lambda_a$ approaches the density $\lambda_u$ of user equipments (UEs) \cite{Hwang}, \cite{AndrewsMag}. In order to exploit the full system resources, traditional frequency reuse schemes are not employed which inevitably results in significant interference that has to be taken carefully into account in system design and performance analysis.

With an increasing network density, the task of optimally placing the APs on the Euclidean plane becomes difficult, if not impossible. Therefore, the  APs will typically have an irregular, random deployment which is expected to affect system performance. Recent research has showed that such randomly deployed systems can be successfully analyzed by employing tools from stochastic geometry \cite{Baccelli}. While significant results have been achieved \cite{Andrews}, \cite{ElSawy}, most of these works (a) assume $\lambda_u$ much larger than $\lambda_a$, and/or (b) ignore the effect of multiple access on performance. This inevitably leads to focusing only on a single resource block and a corresponding analysis on the achieved signal-to-interference (SIR) ratio. Although SIR analysis is certainly very important, e.g., a minimum SIR value is required for UE operation, it provides only partial insight on the achieved user rate \cite{AndrewsMag} as the latter is  affected by the presence of contending UEs sharing the same AP resources.

A few recent works have attempted to address these issues. The random UE distribution is taken into account in \cite{Lee} , \cite{Li}, by incorporating in the analysis the probability of an AP being inactive (no UE present within its cell). However, analysis considers only the SIR (or quantities directly depended upon it). In \cite{Cao}, \cite{Singh}, the achieved user rate is computed assuming time-division-multiple-access (TDMA), however, without taking into account the possibility of inactive APs in the system. In \cite{Dhillon}, \cite{Li2}, performance under a space-division-multiple-access (SDMA) scheme is considered without taking into account TDMA effects. None of these references examines the effect of frequency-division-multiple-access (FDMA) and either the minimum achievable rate corresponding to operating on the SIR threshold is only considered, or the effect of SIR outage on rate is neglected, i.e., rates corresponding to arbitrarily small SIRs are shown to be achievable with a non-zero probability.

In this paper, the stochastic geometry framework is employed to investigate the efficiency of frequency and space dimensions for serving multiple UEs. For this purpose two orthogonal multiple access schemes are considered, the first fully exploiting the available frequency dimensions with spatial dimensions having a secondary role as a means to reduce the effect of TDMA (time sharing), whereas the roles are reversed in the second. The performance measure of interest is the achieved user rate, taking into account the possibility of no service due to SIR outage. Simple, (quasi-) closed form expressions for the user rate distribution are derived that allow for evaluation of performance without the need of computationally intensive simulations. It is shown that performance critically depends on the choice of system parameters such as number of frequency subchannels and transmit antennas. In particular, increasing the number of available frequency dimensions alone is beneficial for users experiencing large interference, whereas increasing spatial dimensions without employing frequency dimensions degrades performance. However, best performance is achieved when both dimensions are combined in serving the users.

The paper is organized as follows. Section II describes the system model with the corresponding SIR and rate analysis performed in Sect. III for the class of non-cooperative multiple access schemes. In Sect. IV, two simple instances of multiple access schemes are described that allow for numerical evaluation of performance, depicted in Sect. V, along with discussion on the selection of system parameters. Section VI concludes the paper.

\section{System Model}
The downlink of an interference-limited, dense wireless network is considered. The randomly deployed over $\mathbb{R}^2$ multiple-antenna APs are modeled as a  homogeneous Poisson point process (PPP) $\Phi_a$ with density $\lambda_a$. The distribution of single-antenna UEs is also assumed to be a realization of another independent PPP $\Phi_u$ with density $\lambda_u$. Each UE is served by its closest AP resulting in non-overlapping, irregular cell shapes that form a Voronoi tessellation of the space \cite{Baccelli}.
For later reference, the probability mass function (pmf) of the number of UEs $K\geq0$ associated with a random AP in the system is given by \cite{Yu}
\begin{equation} \label{eq:Kpmf}
\Pr\{K\}=\frac{3.5^{c}\Gamma(K+c)(\lambda_u/\lambda_a)^K}{\Gamma(c)K!(3.5+\lambda_u/\lambda_a)^{K+c}},
\end{equation}
where $c=3.5$ and $\Gamma(\cdot)$ is the Gamma function. Note that the mean value of $K$ equals $\lambda_u/\lambda_a$ and $\Pr\{K=0\}>0$ for $\lambda_u > 0$, i.e., there is a non-zero probability of a random AP being inactive.

A non-cooperative scenario is considered, where each AP independently serves its associated UEs via a multiple access scheme (to be specified later) utilizing orthogonal time, frequency and spatial dimensions (resources), so as to avoid intra-cell interference. Specifically, the system bandwidth is partitioned into $N$ flat-fading subchannels allowing for parallel transmissions in frequency (FDMA). In addition, each AP is equipped with $M_{\textrm{max}}$ transmit antennas that are used to serve multiple UEs within one time-frequency resource block (SDMA). For analytical purposes, zero-forcing precoding is employed \cite{Dhillon}, with one antenna reserved for each of the $M \leq M_{\textrm{max}}$ concurrently served UEs, i.e., $M$ transmit antennas are active and  $M_{\textrm{max}} - M$ antennas inactive. Note that no beamforming (diversity) gain is considered, even when there are more available antennas than UEs, in order to simplify the problem and focus attention on how efficient the spatial dimensions are in serving multiple UEs. The reader is referred to \cite{Li2} for a first attempt in describing performance when beamforming is also employed in a setting similar to this paper. All active APs transmit with the same, constant over frequency, power that is normalized to unity and allocate equal power percentage to UEs served via SDMA. When the number of UEs served on the same subchannel cannot be supported in one time-frequency resource block via SDMA, time sharing is also employed (TDMA). Note that $M_{\textrm{max}}$ and $N$ are system parameters that are selected offline based on criteria that will be discussed in the following. 

In order to examine system performance with respect to user experience, a typical UE is considered, located  at the origin and served by its closest AP of index, say, 0. Treating interference as noise, the SIR achieved on a considered subchannel is given by (thermal noise is neglected):
\begin{equation} \label{eq:SIR}
\textrm{SIR} = \frac {\frac{1}{M} g_{0} r_0^{-\alpha}}{\sum_{i\in \Phi_{a}\setminus \{0\}}\mathbf{1}\{\tilde{M}_{i} \geq 1\}\frac{1}{\tilde{M}_{i}}g_{i}r_i^{-\alpha}},
\end{equation}
where $r_0\geq0$ is the distance from the serving AP, $\alpha > 2$ the path loss exponent, $M$ is the number of UEs served by AP 0 on the specific subchannel via SDMA, including the typical one ($1 \leq M \leq M_{\textrm{max}}$), and $g_{0}\geq0$ the channel gain from the resulting transmission link. The denominator in (\ref{eq:SIR}) represents the interference power  experienced at the considered subchannel, where $r_i$, $g_{i}$, and $\tilde{M}_{i}$ are the distance, interference channel gain, and number of SDMA-served UEs of AP $i$, respectively ($0 \leq \tilde{M}_{i} \leq M_{\textrm{max}}$). $\mathbf{1}\{\cdot\}$ denotes the indicator function which is employed to account for inactive APs on the considered subchannel, due to absence of UEs in their respective cell and/or resource allocation decisions ($\mathbf{1}\{\tilde{M}_{i} \geq 1\}/\tilde{M}_i \triangleq 0$, for $\tilde{M}_i=0$) . Considering independent identically distributed Rayleigh fading links between APs and UEs, it can be shown that $g_{0} \sim \mathcal{G}(1, 1)$ and $g_{i} \sim \mathcal{G}(\tilde{M}_{i}, 1)$, where $\mathcal{G}(\cdot,\cdot)$ denotes the gamma probability density function (pdf) \cite{Dhillon}. 

Note that conditioning on the existence of a typical UE does not change the distribution of the other UEs in the system\cite{Baccelli}. However, as discussed in \cite{Yu}, the pmf of the number of UEs associated with AP 0 \emph{in addition to} the typical UE and averaged over $\Phi_a$ is given by (\ref{eq:Kpmf}) with $c=4.5$, i.e., there is a slightly larger probability of serving more UEs than any other random AP of the system.

\section{SIR and Rate Distributions}
In this section, the relevant performance measures from the (typical) user perspective and their statistical description are examined.

\subsection{SIR Distribution}
The interference term of (\ref{eq:SIR}) can be viewed as the result of a marked PPP \cite{Baccelli}, with marks $\{(g_i, \tilde{M}_i)\}_{i \neq 0}$. For the case when the marks are mutually independent given the locations of the APs, and the marginal probability density function (pdf) of each mark depends only on the location of its corresponding AP, the statistical characterization of the interference can be obtained by standard methods \cite{Baccelli}. However, this is not the case here since $\{\tilde{M}_i\}$ depend on the number of UEs associated with each AP which, in turn, depend on the position of all APs in the system as the latter determine the area of each cell.

As in \cite{Lee},\cite{Li},\cite{Li2}, the pmf of the number of UEs in a cell given the positions of all APs is approximated here by the pmf of (\ref{eq:Kpmf}) which corresponds to averaging over AP positions. The accuracy of this approximation will be examined by simulations in Sect. V. Under this approximation, $\{\tilde{M}_i\}$ become independent of AP positions, with the corresponding marginal pmf, $\Pr\{\tilde{M}\}$, depending, in addition to the number of UEs in the cell, on the system parameters $N$, $M_{\textrm{max}}$, as well as the multiple access scheme (the irrelevant index $i$ is dropped). The SIR distribution can now be obtained in (quasi-) closed form as follows:
\newtheorem{proposition}{Proposition}
\begin{proposition}
The cumulative distribution function (cdf) of the SIR, $F_{\textrm{SIR}}(\theta) \triangleq \Pr\{\textrm{SIR} \leq \theta \}$, is given by 
\begin{equation} \label{eq:F_SIR}
F_{\textrm{SIR}}(\theta) = \sum_{M = 1}^{M_{\textrm{max}}} \Pr\{M\} F_{\textrm{SIR}}(\theta | M),
\end{equation}
where $\Pr\{M\}$ is the pmf of $M$, determined by the UE distribution and the multiple access scheme, and
\begin{equation} \label{eq:F_SIR_K0}
F_{\textrm{SIR}}(\theta | M) = 1-\frac{1}{1+\rho(\theta,M)},
\end{equation}
is the cdf of the SIR conditioned on $M-1$ additional UEs co-served with the typical UE via SDMA, with
\begin{equation} \label{eq:rho}
\rho(\theta, M) \triangleq \sum_{\tilde{M}=1}^{M_{\textrm{max}}} \Pr\{\tilde{M}\} \theta ^{\frac{2}{a}} \int_{\theta^{-\frac{2}{a}}}^\infty \phi(u^{-\alpha/2} ,M, \tilde{M}) du,
\end{equation}
where $ \phi(x ,y, z) \triangleq 1 - (1 + xy/z)^{-z}$.
\end{proposition}
\begin{IEEEproof}
Eq. (\ref{eq:F_SIR}) is an application of the total probability theorem, whereas (\ref{eq:F_SIR_K0}), (\ref{eq:rho}) are a special case of a more general recent result presented in \cite{Li2} without proof. Direct derivation can be obtained by noticing that the Laplace transform $\mathcal{L}_I(s)$ of the interference, conditioned on $r_0$, is given by a slight modification of \cite[Appendix B]{Dhillon} as
\begin{equation} \label{eq:L_I}
\mathcal{L}_I(s) = \exp \left\{-2 \pi \lambda_a \sum_{\tilde{M}=1}^{M_{\textrm{max}}} \Pr\{\tilde{M} \} \int_{r_0}^{\infty}  \phi(u^{-\alpha} ,s , \tilde{M} ) u du \right\},
\end{equation}
and proceeding as in \cite{Andrews}.
\end{IEEEproof}
Note that Proposition 1 subsumes as special case the scenario considered in \cite{Dhillon} where $\Pr\{M = M_{\textrm{max}}\}=\Pr\{\tilde{M} = M_{\textrm{max}} = 1$, i.e., $M_{\textrm{max}}$ UEs are always available for transmission in every AP of the system. In this case, it can be easily seen that $F_{\textrm{SIR}}(\cdot)$ is a decreasing function of $M_{\textrm{max}}$, mainly due to the decreased transmit power per served UE. This suggests that, at least in terms of SIR performance, an SDMA system may not preferable, as also stated in \cite{Dhillon}. However, this statement should be viewed with caution, as it assumes $\lambda_u \gg \lambda_a$ which does not hold in a dense network deployment and neglects the effect of $N$ and multiple access scheme which, as will be shown, may compensate for the resulting power loss due to SDMA.

\subsection{User Rate Distribution}

Knowledge of (\ref{eq:F_SIR}) is of importance as it provides the probability $F_{\textrm{SIR}}(\theta_0)$ of a service outage due to inability of  UE  operation below the SIR threshold $\theta_0$. In addition, $F_{\textrm{SIR}}(\theta)$ can be used to obtain cdfs of other directly related quantities of interest by transformation of variables. One such quantity employed extensively in the related literature, e.g., \cite{Andrews}, \cite{Lee}, \cite{Li}, \cite{Dhillon}, is the rate $\overline{R}$ achieved \emph{per channel use}, i.e, on a single time-frequency resource block, given by
\begin{equation} \label{eq:R_bound}
\overline{R} \triangleq \log_2(1+\textrm{SIR})/N \textrm{ (b/s/Hz).}
\end{equation}
Examining $\overline{R}$ is important from the viewpoint of system throughput \cite{Li}, \cite{Li2} but provides little insight on the actual achieved user rate. Note that $\overline{R}$ is an upper bound on the actual user rate that can be achieved only when the number of UEs assigned to a subchannel does not exceed $M_{\textrm{max}}$. When this is not the case, the actual rate will necessarily be a fraction of $\overline{R}$ due to time sharing. In an attempt to remedy this issue, $\overline{R}$ is divided by the number of UEs sharing the subchannel in \cite{Cao}, \cite{Singh} (case of $N=M_{\textrm{max}}=1$ is only considered). However, this may still be a misleading measure of performance as it does not take into account the probability of an SIR outage and, therefore, provides overconfident results in the small rate region.

In order to avoid these issues, the achieved rate of a typical UE is defined in this paper as
\begin{equation} \label{eq:R}
R \triangleq \mathbf{1}\{\textrm{SIR} \geq \theta_0\} \overline{R} M/K_0 \textrm{ (b/s/Hz)},
\end{equation}
where $K_0\geq 1$ denotes the number of UEs, including the typical one, assigned to the considered subchannel and $M = \min\{K_0, M_{\textrm{max}}\}$ is the number of UEs served via SDMA in each resource block. Note that (\ref{eq:R}) implies that a registered UE in the system will be allocated part of the available resources, and thus affect the achieved rate of other UEs, irrespective of being unable to decode data due to SIR outage. This corresponds to the realistic assumption that system registration and signaling transmission requires a much lower SIR threshold than data transmission. The following proposition holds for the rate distribution:

\begin{proposition}
The cdf of $R$ is given by
\begin{equation} \label{eq:F_R}
F_{R}(r) = \sum_{K_0 \geq 0} \Pr\{K_0\} F_{R}(r|K_0),
\end{equation}
where $\Pr\{K_0\}$ is the pmf of $K_0$, determined by the UE distribution and the multiple access scheme, and $F_{R}(r|K_0)$ is the rate cdf conditioned on $K_0-1$ additional UEs co-served with the typical UE on the same subchannel, given by
\begin{equation} \label{eq:F_R_K0}
F_{R}(r|K_0)=\left\{\begin{IEEEeqnarraybox}[\relax][c]{l's}
 F_{\textrm{SIR}}(\theta_0|M),&$r \leq \frac{\overline{R}_{1,\theta_0}M}{N K_0},$ \\
F_{\textrm{SIR}}\left(2^{rNK_0/M}-1|M\right),&$r \geq \frac{\overline{R}_{1,\theta_0}M}{NK_0},$
\end{IEEEeqnarraybox}\right.
\end{equation}
where $ F_{\textrm{SIR}}(\theta|M)$ is given in (\ref{eq:F_SIR_K0}) and ${\overline{R}_{1,\theta_0}} \triangleq \log_2(1+\theta_0)$.
\end{proposition}
\begin{IEEEproof}
Eq. (\ref{eq:F_R}) is an application of total probability theorem. Denoting the SIR outage event $\{\textrm{SIR} < \theta_0\}$ and its complement, $\{\textrm{SIR} \geq \theta_0\}$, as $\mathcal{O}$ and ${\mathcal{O}}^c$, respectively, it follows from (\ref{eq:R}) that $F_{R}(r|K_0)$ can be written as
\begin{eqnarray}
F_{R}(r|K_0)&{}={}& F_{\textrm{SIR}}(\theta_0) F_{R}(r|K_0,\mathcal{O})\nonumber\\
&&{+}\:(1-F_{\textrm{SIR}}(\theta_0)) F_{R}(r|K_0,{\mathcal{O}}^c).
\end{eqnarray}
Conditioned on $\mathcal{O}$, $R=0$, therefore, $F_{R}(r|K_0,\mathcal{O}) = 1, \forall r, K_0$, whereas, conditioned on $\mathcal{O}^c$,
\begin{eqnarray}
F_{R}(r|K_0,{\mathcal{O}}^c) &{}={}& F_{\textrm{SIR}}(\tilde{\theta}|{\mathcal{O}}^c)\nonumber\\
&{}={}& \left\{\begin{IEEEeqnarraybox}[\relax][c]{l's}
\frac{F_{\textrm{SIR}}(\tilde{\theta})-F_{\textrm{SIR}}(\theta_0)}{1-F_{\textrm{SIR}}(\theta_0)},&$\tilde{\theta} \geq \theta_0,$\\
0, &$\tilde{\theta} < \theta_0,$
\end{IEEEeqnarraybox}\right.
\end{eqnarray}
where $\tilde{\theta} \triangleq 2^{rNK_0/M}-1$ and the last equality follows from basic probability theory and the continuity of $F_{\textrm{SIR}}(\theta)$. Combining (11) and (12) gives (\ref{eq:F_R_K0}).
\end{IEEEproof}

Note that the upper term of (\ref{eq:F_R_K0}) indicates that for small values of $r$, rate outage probability coincides with the SIR outage probability, irrespective of the actual value of $r$, since it is the SIR outage event (strong interference) that prevents UEs from achieving small rates. In contrast, for larger rates, i.e., large operational SIR, performance is also limited by the effect of multiple access.

\section{Multiple Access Schemes}
The performance measures of the previous section depend on the statistics of $K_0$, $M$ and $\tilde{M}$, which, in turn, are determined by the multiple access scheme employed and the system parameters $N$ and $M_{\textrm{max}}$. In order to investigate the efficiency of frequency and spatial dimensions in serving multiple UEs in an interference-limited network, two simple schemes are presented, one relying more heavily on the FDMA capabilities of the system, the other on SDMA. Their performance will be investigated in Sect. V.

\subsection{Scheme 1}
This scheme fully exploits frequency dimensions by first assigning UEs evenly to the available subchannels according to the following procedure ($ \lfloor \cdot \rfloor$ is the integer floor operator):

\begin{algorithm}
\caption{Subchannel Allocation}
\begin{algorithmic}[1]
\STATE Define $\mathcal{N} \subseteq \{1,2,\ldots,N\}$ the set of available subchannels for allocation at any given instant.
\STATE Randomly order the $K$ cell users via an index $k \in \{1,2,\ldots,K\}$.
\FOR{$L=0$ to $\lfloor K/N \rfloor$} 
\STATE $\mathcal{N} \gets \{1,2,\ldots,N\}$;
\FOR{$k=LN+1$ to $\min\{LN+N,K\}$} 
\STATE Assign UE $k$ a random subchannel $n_k \in \mathcal{N}$; 
\STATE $\mathcal{N} \gets \mathcal{N}\backslash\{n_k\}$;
\ENDFOR
\ENDFOR
\end{algorithmic}
\end{algorithm}

After subchannel allocation has been determined, transmission to multiple UEs within a single subchannel is performed via SDMA in case the number of UEs does not exceed $M_{\textrm{max}}$, whereas the case of more UEs is handled by a combination of SDMA and TDMA.

Note two features of the subchannel allocation process. First, subchannel allocation is implemented in a fair but random fashion, i.e., no channel information is taken into account, with the corresponding results serving as a lower bound on performance for a more intelligent scheme. Second, only one subchannel is allocated to each UE which may lead to a reduced bandwidth utilization, i.e., unused subchannels, in case of the number of subchannels being greater than the number of UEs. This should not be viewed as a serious issue, since a) $N$ is a flexible parameter that can be set to a small enough value, e.g., $N=1$, so as to avoid this case, if desired, and b) presence of unused subchannels may be beneficial as it results in reduced interference.

The probability of a random AP transmitting on a certain subchannel is computed in the appendix to be  $\Pr\{\tilde{M} > 0\} = \sum_{K\geq0}\Pr\{K\} \min\{K,N\}/N$, which shows that increasing $N$ increases the SIR since the terms $\Pr\{\tilde{M}\},\tilde{M}>0$, in (\ref{eq:rho}) decrease accordingly. However, there exists a trade-off since, as it can be easily verified, $R\rightarrow 0$ for $N \gg 1$, i.e., SIR increase comes at a cost of smaller rate due to small bandwidth utilization per subchannel.

\subsection{Scheme 2}
This scheme fully exploits the spatial dimensions by first randomly constructing groups of $M_{\textrm{max}}$ UEs that will be served via SDMA (note that the last constructed group may consist of less UEs). These groups are subsequently allocated subchannels according to the same principle as Scheme 1. TDMA is employed when more than one SDMA groups are served by the same subchannel. Typical realizations of Schemes 1 and 2 are shown in Fig. 1.

The rationale of Scheme 2 is to further reduce subchannel activity. In particular, it can be shown that $\Pr\{\tilde{M} > 0\} = \sum_{K\geq0}\Pr\{K\} \min\{\left\lceil K/M_{\textrm{max}} \right\rceil,N\}/N$, where $\left\lceil \cdot \right\rceil$ denotes the integer ceiling operator, which is strictly smaller than the corresponding one of Scheme 1 for the same $N$ and $M_{\textrm{max}}>1$. However, the resulting interference reduction comes at a cost of reduced transmit power to each SDMA-served UE.

In the following section, performance under these schemes is investigated in order to obtain insights on how the available frequency and/or spatial degrees should be incorporated by the system.

\begin{figure}
\centering
\includegraphics[width=8.5cm]{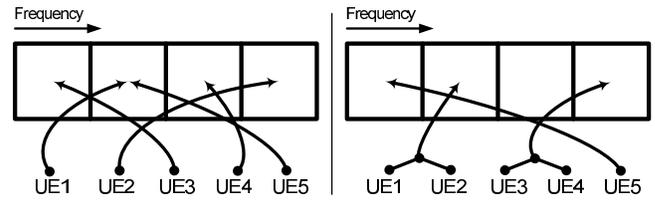}
\caption{Typical realizations of subchannel allocation for Scheme 1 (left) and Scheme 2 (right), with $K=5$, $N=4$, $M_{\textrm{max}}=2$. SDMA is employed for UE pair $(1,5)$ in Scheme 1 and UE pairs $(1,2)$, $(3,4)$ in Scheme 2. One subchannel is inactive with Scheme 2.}
\end{figure}

\section{Numerical Results and Discussion}
This section examines the performance of the network in terms of user rate $R$ for the two multiple access schemes described in the previous section. Note that the analysis of Sec. III allows for relatively simple numerical evaluation of performance without the need of computationally intensive simulations. However, it is necessary to provide the pmfs for $K_0$, $M$ and $\tilde{M}$, under each scheme. Although evaluation can in principle be done analytically, the corresponding results do not provide any insight and are omitted for brevity. In any case, they can be computed relatively easily via Monte Carlo simulation.

In the following, a test case scenario with $\lambda_u/\lambda_a=10$ is considered unless stated otherwise. Note that this case corresponds to an average of 10 UEs associated with a random AP with the probability of having more than 40 UEs in a cell being practically zero. The path loss exponent is set to $\alpha =4$ and SIR threshold is set to $\theta_0 = 0$ dB.

\begin{figure}
\centering
\psfragscanon
\psfrag{x12}[t][t]{$0$}
\psfrag{x13}[t][t]{$0.05$}
\psfrag{x14}[t][t]{$0.1$}
\psfrag{x15}[t][t]{$0.15$}
\psfrag{x16}[t][t]{$0.2$}
\psfrag{x17}[t][t]{$0.25$}
\psfrag{x18}[t][t]{$0.3$}
\psfrag{x19}[t][t]{$0.35$}
\psfrag{x20}[t][t]{$0.4$}
\psfrag{x21}[t][t]{$0.45$}
\psfrag{x22}[t][t]{$0.5$}

\psfrag{v12}[r][r]{$0$}
\psfrag{v13}[r][r]{$0.1$}
\psfrag{v14}[r][r]{$0.2$}
\psfrag{v15}[r][r]{$0.3$}
\psfrag{v16}[r][r]{$0.4$}
\psfrag{v17}[r][r]{$0.5$}
\psfrag{v18}[r][r]{$0.6$}
\psfrag{v19}[r][r]{$0.7$}
\psfrag{v20}[r][r]{$0.8$}
\psfrag{v21}[r][r]{$0.9$}
\psfrag{v22}[r][r]{$1$}

\psfrag{s01}[c][c]{$r$ (b/s/Hz)}
\psfrag{s02}[c][c]{$F_{R}(r)$}
\psfrag{s30}[c][c]{$N = 5$}
\psfrag{s29}[l][l]{$N = 50$}
\psfrag{s28}[l][c]{Scheme 1}
\psfrag{s39}[l][c]{Scheme 2}
\psfrag{s21}[l][c]{$M_{\textrm{max}} = 1$}
\psfrag{s22}[l][c]{$M_{\textrm{max}} = 3$}
\psfrag{s23}[l][c]{$M_{\textrm{max}} = 5$}

\psfrag{s06}[t][t][0]{}
\psfrag{s05}[t][t][0]{}

\resizebox{8.8cm}{!}{\includegraphics{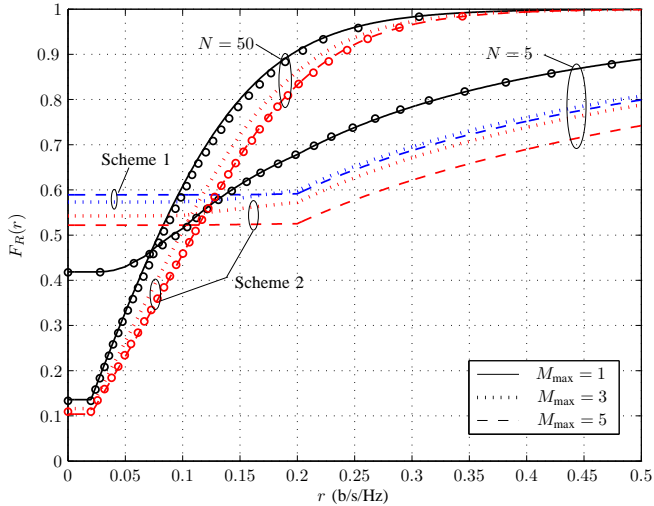}}
\caption{$F_R(r)$ of Schemes 1 and 2 for various values of $N$ and $M_{\textrm{max}}$ ($\lambda_u/\lambda_a = 10$, $\alpha = 4$, $\theta_0 = 0$ dB). Circles depict simulations results.}
\end{figure}

\emph{1) Exploiting the frequency dimensions:} Figure 2 depicts the analytically evaluated $F_R(r)$ based on (\ref{eq:F_R}) under different combinations of system parameters $N=5, 50$, and $M_{\textrm{max}}=1,3,5$. Simulation results are also shown (circles) for the cases of $(M_{\textrm{max}},N)=(1,5),(1,50),(5,50)$.  It can be seen that there is a good correspondence between analytical and simulated results justifying the validity of the approximations employed in Sect. III.A. Simulation results for the other cases show the same behavior and are omitted for clarity.

In order to examine the efficiency of exploiting frequency dimensions alone ($M_{\textrm{max}}=1$), the values of $N=5, 50$ were selected as examples of cases when the number of subchannels is smaller  and larger than $\lambda_u/\lambda_a$, respectively. The first case results in having all subchannels occupied with high probability (large system bandwidth utilization), with each of them typically serving more than one UE via TDMA. In the second case, the number of subchannels exceeds the number of UEs with high probability, therefore, there will be unoccupied subchannels and TDMA is not required for any active subchannel. As can be seen from the corresponding curves, a small $N$ provides better performance when high rates are considered (center-cell UEs). However, this comes at a cost of a high SIR outage and a correspondingly unfair system performance, with the largest percentage of UEs not supported at all. In contrast, increasing $N$ results in a more fair system performance, where smaller rates can be supported with a lower outage probability, at a cost of reducing the system capability of providing higher rates. 

\emph{2) Exploiting the spatial dimensions:} Consider first the case when $N=5$. Under Scheme 1, incorporating the available spatial dimensions results in improving the performance for higher rates, but at the same time degrades performance for lower rates. This can be explained by noting that employing more than one transmit antennas under Scheme 1 does not affect interference significantly since subchannel activity remains the same. However, the reduced transmit power per UE provided by the scheme severely affects UEs experiencing large interference (lower rate region), whereas in the higher rate region, corresponding to UEs experiencing low interference, performance benefits from reducing the time-sharing effect. Note that performance of Scheme 1 for $M_{\textrm{max}} = 5$ is very close to $M_{\textrm{max}} = 3$ since there is a small probability of having more than 3 UEs allocated to the same subchannel. Scheme 2 shows a better performance that actually improves with increasing $M_{\textrm{max}}$. Clearly, employing the spatial dimensions as a means to further reduce subchannel activity is beneficial and partially compensates for the loss imposed by the transmit power reduction. However, both schemes are worse than the case of $M_{\textrm{max}}=1$ in the low rate region.

For the case of $N=50$, performance of Scheme 1 is not shown for clarity as it is virtually identical with that of $M_{\textrm{max}}=1$ due to the fact that there is a very small probability of having more than one UE associated with a subchannel. On the other hand, Scheme 2 manages to improve performance as in the case of $N=5$, this time even for the lower rate region. This is because the large number of subchannels alone guarantees a high operational SIR and transmit power reduction due to SDMA has little effect that is more than compensated by the further interference reduction Scheme 2 offers.

\begin{figure}
\centering
\psfragscanon
\psfrag{x12}[t][t]{$0$}
\psfrag{x13}[t][t]{$0.1$}
\psfrag{x14}[t][t]{$0.2$}
\psfrag{x15}[t][t]{$0.3$}
\psfrag{x16}[t][t]{$0.4$}
\psfrag{x17}[t][t]{$0.5$}
\psfrag{x18}[t][t]{$0.6$}
\psfrag{x19}[t][t]{$0.7$}
\psfrag{x20}[t][t]{$0.8$}
\psfrag{x21}[t][t]{$0.9$}
\psfrag{x22}[t][t]{$1$}

\psfrag{v12}[r][r]{$0$}
\psfrag{v13}[r][r]{$0.1$}
\psfrag{v14}[r][r]{$0.2$}
\psfrag{v15}[r][r]{$0.3$}
\psfrag{v16}[r][r]{$0.4$}
\psfrag{v17}[r][r]{$0.5$}
\psfrag{v18}[r][r]{$0.6$}
\psfrag{v19}[r][r]{$0.7$}
\psfrag{v20}[r][r]{$0.8$}
\psfrag{v21}[r][r]{$0.9$}
\psfrag{v22}[r][r]{$1$}

\psfrag{s13}[l][l]{$M_{\textrm{max}}=5$, $N=1$}
\psfrag{s14}[l][l]{$M_{\textrm{max}}=1$, $N=1$}
\psfrag{s15}[l][l]{$M_{\textrm{max}}=1$, $N=N^*(r_0)$}
\psfrag{s16}[l][l]{$M_{\textrm{max}}=5$, $N=N^*(r_0)$, Scheme 1}
\psfrag{s17}[l][l]{$M_{\textrm{max}}=5$, $N=N^*(r_0)$, Scheme 2}

\psfrag{s34}[l][l]{$\lambda_u/\lambda_a = 10$}
\psfrag{s43}[l][l]{$\lambda_u/\lambda_a = 2$}

\psfrag{s01}[c][c]{$r_0$ (b/s/Hz)}
\psfrag{s02}[c][c]{rate outage probability}

\psfrag{s06}[t][t][0]{}
\psfrag{s07}[t][t][0]{}

\resizebox{8.8cm}{!}{\includegraphics{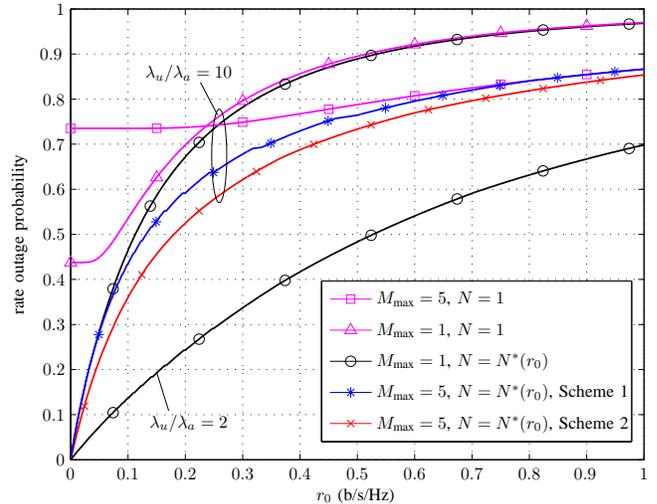}}
\caption{Rate outage probability with optimal selection of $N$ ( $\alpha = 4$, $\theta_0 = 0$ dB).}
\end{figure}

\emph{3) Optimal selection of $N$:} From the results of Fig. 2 it is clear that system performance is highly dependent on the selection of $N$ and $M_{\textrm{max}}$, whose optimal value varies with rate. It is desirable to optimize these parameters for a given rate $r_0$ of interest, i.e., minimize rate outage probability $F_R(r_0)$, and the analytical formulas of Sec. III allow for a relatively simple numerical search. An example of such a procedure is shown in Fig. 3 where rate outage probability is depicted, for fixed $M_{\textrm{max}}$ (equal to $1$ or $5$), under the optimal choice of the number of subchannels $N^*(r_0)$. Note that the curves corresponding to the optimal $N^*(r_0)$ should not be interpreted as cdf plots since $N$ varies with $r_0$. They only inform the system designer of the minimum possible rate outage probability for a certain rate $r_0$. This minimum is achieved by a value $N^*(r_0)$ which is not necessarily optimal for other rates. Performance under the commonly considered case of $N=1$ (no FDMA) is also depicted for reference. 

It can be seen that with no SDMA ($M_{\textrm{max}}=1$) optimal selection of $N$ results in improved performance compared to the $N=1$ case, with notable difference in the small rate region. When  SDMA is employed ($M_{\textrm{max}}=5$) without FDMA, performance is severely degraded in the low rate/SIR region, making SDMA not a viable option in this case as was also observed in \cite{Dhillon}. However, conclusions are different when SDMA is employed along with FDMA as both Schemes 1 and 2 provide better performance. Notice how Scheme 1 coincides with the $M_{\textrm{max}} =1, N=N^*(r)$ case in the low rate region since the number of subchannels is so large that there is no use for SDMA, and with the $M_{\textrm{max}} =5,N=1$ case in the high rate region since $N=1$ is the optimal choice there. Scheme 2 uniformly outperforms Scheme 1 for the depicted $r$ range (Scheme 1 coincides with Scheme 2 at $r=1.7$ where the optimal $N$ becomes 1). These results clearly show that SDMA is a viable option for an interference-limited network, \emph{as long as it is efficiently combined with FDMA}.

\emph{4) Effect of network densification:} In order to investigate the trend of network densification under a fixed UE distribution, the performance for $\lambda_u/\lambda_a = 2$ is also shown in Fig. 2 with $M_{\textrm{max}} =1, N=N^*(r)$ . Note that in this case the average number of transmit antennas per unit area is the same as the case $\lambda_u/\lambda_a = 10$ with $M_{\textrm{max}} =5$, although they are not directly comparable since there are practical differences in terms of hardware, deployment cost, e.t.c. As can be seen, densification of the network increases performance by a considerable margin even when a single transmit antenna is employed.

\section{Conclusion}
In this paper the efficiency of frequency and spatial dimensions in serving multiple UEs in a small cell network was examined. A simple expression for the achieved user rate was provided, taking into account the effects of SIR outage and multiple access. Performance was investigated for two simple non-cooperative multiple access schemes. It was shown that appropriate selection of system parameters and multiple access scheme can provide large performance gains for an operational region of interest. FDMA was shown to effectively reduce interference which is of essential importance in the type of network considered. SDMA can provide additional performance gains, however, only when combined with FDMA.

\appendices
\section{Computation of $\Pr\{\tilde{M} > 0\}$ for Scheme 1}
By symmetry of the subchannel allocation procedure, the following proof holds for any subchannel. Consider a random \newpage \noindent AP associated with $K$ randomly ordered UEs and let  $\Pr\{\tilde{M} > 0|K\}$ denote the probability of assigning at least one UE on  subchannel, say, 1. Clearly, $\Pr\{\tilde{M} > 0|K\}=1$ for $K \geq N$ and $\Pr\{\tilde{M} > 0|K\}=0$ for $K=0$. For the case $0<K<N$, define the mutually exclusive events $\mathcal{A}_m \triangleq$ $\{$$m$-th UE is assigned subchannel 1$\}, m=1,2,\ldots, K$. It is easy to see that \
\begin{equation} \label{eq:Pr_Bm}
\Pr\{\mathcal{A}_m\} = \frac{1}{N-(m-1)} \prod_{r=1}^{m-1}\left( 1 - \frac{1}{N-(r-1)}\right),
\end{equation}
and
\begin{equation} \label{eq:p_N_K}
\Pr\{\tilde{M} > 0|K\} = \sum_{m=1}^{K} \Pr\{\mathcal{A}_m\} = K/N, 0<K<N,
\end{equation}
where the last equality follows by simple algebra. Averaging $\Pr\{\tilde{M} > 0|K\}$ over $K$ gives $\Pr\{\tilde{M} > 0\}$.

\section*{Acknowledgment}
This work has been performed in the context of the ART-COMP PE7(396)\textit{ ``Advanced Radio Access Techniques for Next Generation Cellular NetwOrks: The Multi-Site Coordination Paradigm''}, THALES-INTENTION and THALES-ENDECON research projects, within the framework of Operational Program ``Education and Lifelong earning'', co-financed by the European Social Fund (ESF) and the Greek State.



%

\end{document}